\documentclass[aps, prd, amsmath, floats, floatfix, twocolumn, nofootinbib, showpacs]{revtex4-1}
\usepackage{graphicx}
\usepackage{color}

\newcommand{\be}{\begin{eqnarray}}
\newcommand{\ee}{\end{eqnarray}}

\begin{document}

\title{Constraining the Cardoso-Pani-Rico metric with future observations of SgrA$^*$}

\author{Cosimo Bambi}
\email{bambi@fudan.edu.cn}

\affiliation{Center for Field Theory and Particle Physics and Department of Physics, 
Fudan University, 200433 Shanghai, China}

\date{\today}

\begin{abstract}
SgrA$^*$, the supermassive black hole candidate at the center of our Galaxy, 
seems to be one of the most promising object to test the Kerr black hole hypothesis with
near future observations. In a few years, it will hopefully be possible to measure 
a number of relativistic effects around this body, and the combination of different 
observations can be used to constrain possible deviations from the Kerr solution. 
In this paper, I discuss the combination of three promising techniques in the 
framework of the Cardoso-Pani-Rico parametrization: the observation of blobs of 
plasma orbiting near the innermost stable circular orbit, the detection of the black 
hole shadow, and timing observations of a radio pulsar in a compact orbit. The 
observations of blobs of plasma and of the shadow can probe the strong gravitational 
field around SgrA$^*$, while the radio pulsar would be sensitive to the weak field 
region at larger radii. In the case of a fast-rotating object, the combination of the 
three measurements could provide strong constraints on the actual nature of SgrA$^*$. 
For a non-rotating or slow-rotating object, the bounds would be weak.
\end{abstract}


\maketitle


\section{Introduction}

Astrophysical black hole (BH) candidates are thought to be the Kerr BHs of 
general relativity, but the actual nature of these objects has still to be confirmed. 
Robust dynamical measurements of their masses clearly indicate that these bodies
cannot be explained in the framework of conventional physics as compact stars 
or clusters of compact stars~\cite{bh}, but the observational evidence that the geometry 
of the spacetime around them is described by the Kerr solution is still lacking. In 
the last few years, there have been significant work to study the possibility of 
testing the Kerr BH hypothesis with present and future observations~\cite{review}. 
Current X-ray data can already rule out some alternative candidates, namely some
exotic compact stars without an event horizon~\cite{iron-bstar} and some kind of 
wormholes~\cite{iron-wh}. However, other candidates like non-Kerr BHs in 
alternative theories of gravity are very difficult to distinguish from the Kerr BHs 
of general relativity~\cite{cfm-iron}.

The Kerr BH hypothesis can be potentially tested by studying the properties of 
the electromagnetic radiation emitted by the gas of the accretion disk. Strictly 
speaking, such an approach can only test the Kerr geometry and measure possible 
deviations from it. It cannot test the Einstein equations, as the Kerr metric is a BH 
solution even in some alternative theories of gravity~\cite{kerr} and the electromagnetic 
spectrum of the accretion disk only depends on the background geometry. 
Bearing this in mind, the ideal strategy would be to use an approach similar to the 
PPN formalism of the weak field regime~\cite{will}, where the most general background 
metric is characterized by a number of parameters to be measured from observations, 
and {\it a posteriori} one can verify if astrophysical data require that the values of 
these parameters is consistent with the predictions of general relativity. The 
observation of a spectrum that looks like the one of a Kerr BH with a certain spin is 
not enough to confirm the Kerr BH hypothesis, because there is typically a degeneracy 
between the measurement of the spin and possible deviations from the Kerr solution, 
with the result that a non-Kerr object may mimic a Kerr BH with a different spin~\cite{cfm-iron}.

Unfortunately, at present there is no general and satisfactory formalism as the PPN 
one to test the Kerr nature of BH candidates. In the strong field regime, an expansion 
in $M/r$ does not work and deviations from the Kerr background can easily generate 
pathological features in the spacetime. A first attempt in this direction was proposed 
in Ref.~\cite{jp-m}, in which the background metric was characterized by a set of an 
infinite number of deformation parameters $\{ \epsilon_k \}$ that measure possible 
deviations from the Kerr solution. Recently, Cardoso, Pani and Rico have proposed an 
extension of the Johannsen-Psaltis metric of Ref.~\cite{jp-m}, in which there are two  
sets of deformation parameters, namely $\{\epsilon^t_k\}$ and $\{\epsilon^r_k\}$~\cite{cpr-m}. 
The Johannsen-Psaltis parametrization is recovered when $\epsilon^t_k = \epsilon^r_k$, 
while the metric reduces to the Kerr solution when $\epsilon^t_k = \epsilon^r_k = 0$. 
As shown in~\cite{cb-cpr}, the Cardoso-Pani-Rico parametrization can describe 
non-Kerr objects with qualitatively different features with respect to the Johannsen-Psaltis one
and it is therefore a more suitable framework to test the Kerr nature of BH candidates.

For the time being, the spacetime geometry around BH candidates can be probed 
with available X-ray data, by studying the thermal spectrum of thin accretion disks~\cite{t-cfm} 
or the shape of the iron K$\alpha$ line~\cite{t-iron}. However, there is a strong 
correlation between the measurements of the spin and of possible deviations from the 
Kerr solution, with the result that a single measurement can only constrain a combination 
of the spin and the deformation parameter~\cite{lingyao,cb-cpr}. In order to really 
test the Kerr BH hypothesis it would be necessary to combine several measurements 
of the same object. Unfortunately, the measurements of the disk's thermal spectrum 
and of the iron line are very sensitive to the same feature, namely the position of the 
innermost stable circular orbit (ISCO), and only with very high quality data, not available 
today, it may be possible to break the degeneracy between the estimates of the spin 
and of the deformation parameter~\cite{cfm-iron}. Other approaches, like the study of 
QPOs~\cite{t-qpo} or the estimate of the jet power~\cite{t-jet}, are not yet mature 
techniques, and therefore they cannot yet be used to test fundamental physics.

In Ref.~\cite{cb-cpr}, it was studied how to constrain the Cardoso-Pani-Rico deformation
parameters from X-ray observations of stellar-mass BH candidates with the analysis of 
the thermal spectrum of thin accretion disks. These data are already available and 
constraints were obtained for two specific sources, namely A0620 and Cygnus X-1. 
The conclusion was that there is a fundamental degeneracy between the measurement 
of the spin and the deformation parameters and therefore we cannot put any bound on 
possible deviations from the Kerr solution. The aim of the present work is to study the 
possibility of constraining the values of these deformation parameters with future 
observations of SgrA$^*$, the supermassive BH candidate at the center of our Galaxy. 
For the time being, there are no observational data to probe the spacetime geometry 
around this object and no reliable approaches to estimate its spin. However, the situation 
could soon change, and SgrA$^*$ may become one of the best candidates to test the 
Kerr BH hypothesis. There is indeed the realistic possibility that it will be soon possible 
to test the geometry around this object with a number of different observations.
The key-point is that such different observations could measure the relativistic effects 
at different radii, which would break the degeneracy between the estimate of the spin 
and of the deformation parameters.
In particular, the instrument GRAVITY may allow to observe blobs of plasma around 
SgrA$^*$ within a few years~\cite{grav,grav2}. Experiments like The Event Horizon 
Telescope may observe the shadow of SgrA$^*$ and thus measure its apparent 
photon capture sphere~\cite{eht1,eht2}. The possible discovery of a radio pulsar in a 
compact orbit around SgrA$^*$ would permit precise measurements of its mass 
and spin, independently of the actual nature of this object~\cite{pulsar}. The 
combination of all these measurements can potentially test the nature of SgrA$^*$,
which is not the case for stellar-mass BH candidates.

The content of the paper is as follows. In Section~\ref{s-2}, I briefly review the 
Cardoso-Pani-Rico parametrization, namely the theoretical framework that will be 
used here to study how to test the Kerr nature of SgrA$^*$, and the possible
future observations of blobs of plasma, BH shadow, and pulsar in a compact orbit. 
Section~\ref{s-3} discusses the constraints from these measurements. Summary 
and conclusions are reported in Section~\ref{s-4}. Throughout the Paper, I use 
units in which $G_{\rm N} = c = 1$.

\section{Testing the nature of S\lowercase{gr}A$^*$ \label{s-2}}

\subsection{Theoretical framework}

In Boyer-Lindquist coordinates, the Cardoso-Pani-Rico parametrization reads~\cite{cpr-m}
\begin{widetext}
\be\label{eq-m}
ds^2 &=& - \left(1 - \frac{2 M r}{\Sigma}\right)\left(1 + h^t\right) dt^2
- 2 a \sin^2\theta \left[\sqrt{\left(1 + h^t\right)\left(1 + h^r\right)} 
- \left(1 - \frac{2 M r}{\Sigma}\right)\left(1 + h^t\right)\right] dt d\phi \nonumber\\ 
&& + \frac{\Sigma \left(1 + h^r\right)}{\Delta + h^r a^2 \sin^2\theta} dr^2
+ \Sigma d\theta^2 
+ \sin^2\theta \left\{\Sigma + a^2 \sin^2\theta \left[ 2 \sqrt{\left(1 + h^t\right)
\left(1 + h^r\right)} - \left(1 - \frac{2 M r}{\Sigma}\right)
\left(1 + h^t\right)\right]\right\} d\phi^2 \, , \quad
\ee
where $M$ is the BH mass, $a = J/M$ is the BH spin parameter, $J$ is the BH spin 
angular momentum, $\Sigma = r^2 + a^2 \cos^2 \theta$, $\Delta = r^2 - 2 M r + a^2$, and
\be
h^t = \sum_{k=0}^{+\infty} \left(\epsilon_{2k}^t 
+ \epsilon_{2k+1}^t \frac{M r}{\Sigma}\right)\left(\frac{M^2}{\Sigma}
\right)^k\, , 
\qquad
h^r = \sum_{k=0}^{+\infty} \left(\epsilon_{2k}^r
+ \epsilon_{2k+1}^r \frac{M r}{\Sigma}\right)\left(\frac{M^2}{\Sigma}
\right)^k \, .
\ee
\end{widetext}
There are two infinite sets of deformation parameters, $\{\epsilon_k^t\}$ and 
$\{\epsilon_k^r\}$. The line element in~(\ref{eq-m}) reduces to the Johannsen-Psaltis
one for $h^t = h^r$, and to the Kerr line element when $h^t = h^r = 0$. In the
Johannsen-Psaltis background, $\epsilon_0 = 0$ to have an asymptotically flat
spacetime, while $\epsilon_1$ and $\epsilon_2$ must be small to meet the 
Solar System constraints~\cite{cpr-m}. $\epsilon_3$ is the first unbounded 
deformation parameter and there are no qualitative differences between 
$\epsilon_3$ and higher order terms~\cite{agn}. In the Cardoso-Pani-Rico 
background, the first unconstrained deformation parameters are $\epsilon^t_3$ 
and $\epsilon^r_2$~\cite{cpr-m}.

The Cardoso-Pani-Rico metric is not a solution of any known alternative 
theory of gravity, but it is an attempt to quantify generic deviations from the Kerr 
solutions through its deformation parameters $\{\epsilon_k^t\}$ and $\{\epsilon_k^r\}$. 
This is the same approach as the PPN metric for Solar System experiments, in which 
the deformation parameters are free and to be determined by observations. Only
measurements of the Schwarzschild and Kerr metrics can constrain these 
deformation parameters, as, in absence of a theory, it is not possible to relate 
them to bounds inferred in other contexts, like the expansion of the Universe or 
the emission of gravitational waves. The Cardoso-Pani-Rico metric is obtained
by considering a generic deformation in the non-rotating Schwarzschild solution 
and then it is performed a Newman-Janis transformation to get a rotating BH. 
With such a procedure, $g_{\theta\theta}$ is not altered and all the deviations
are encoded into two functions, namely $h^t$ and $h^r$, but this does not mean 
that the result is the most general rotating BH solution -- actually, we know it is not. 
As already mentioned, at present there is no satisfactory formalism like the PPN 
one for the strong field regime. The physical implications of this limitation is that 
we cannot really take into account generic deviations from the Kerr metric and 
the line element in~(\ref{eq-m}) may miss some important theoretical motivated 
BH solutions.

\subsection{Orbiting hot-spots}

General relativistic magneto-hydrodynamic simulations of accretion flows 
onto BHs indicate that temporary clumps of matter may be common in the 
region near the ISCO~\cite{grmhd}. SgrA$^*$ exhibits powerful flares in the 
X-ray, NIR, and submillimeter bands, see e.g.~\cite{flares}. A flare typically 
lasts 1-3 hr and shows a quasi-periodic substructure with a time scale of 
about 20 minutes. While the exact mechanism responsible for these flares 
is not known, current data seem to favor the hot spot model, namely 
a blob of plasma orbiting near the ISCO radius~\cite{trippe}. Within a couple
of years, the GRAVITY instrument for the ESO Very Large Telescope 
Interferometer is expected to test the actual nature of these flares and 
hopefully confirm the hot spot model~\cite{grav,grav2}.

Light curves, centroid tracks, and direct images of hot spots orbiting around
SgrA$^*$ can potentially carry a lot of information about the spacetime geometry
around this object, but the situation with real data is much more complicated 
and, except in the case of substantial differences from the Kerr spacetime, 
eventually we can at most get an estimate of the Keplerian frequency at the ISCO 
radius~\cite{zilong} (but see~\cite{n-dan}). The latter does depend on the spin parameter and on 
possible deviations from the Kerr solution and its measurement can thus
constrain the nature of SgrA$^*$ on the spin
parameter--deformation parameter plane.

\subsection{Black hole shadow}

The direct image of a BH surrounded by an optically thin accretion flow is 
characterized by a dark area over a bright background. Such a dark area 
is usually refereed to as the shadow of the BH~\cite{sh0}. While the intensity 
map of the image depends on the exact accretion model and emission mechanisms, 
the shape of the shadow is only determined by the metric of the spacetime 
and corresponds to the boundary of the photon capture sphere as seen by a 
distant observer. The observation of the shadow of a BH can thus be used to 
test the nature of the compact object~\cite{n-brane,sh1}, and shadows of 
many non-Kerr BHs have been calculated~\cite{sh2,sh3,sh4,sh5}.

SgrA$^*$ is the best candidate to observe for the first time the shadow of a 
BH, because it is the BH candidate with the largest angular size on the sky. 
At submillimeter wavelengths, its accretion flow is supposed to become optically 
thin and the interstellar scattering should be significantly lower~\cite{sh0}. 
At first approximation, the shape of the shadow is a circle, whose size is 
essentially set by the mass of the compact object and its distance from us. 
In the case of SgrA$^*$, the radius of the shadow should be about 25~$\mu$as
and not very sensitive to the exact background geometry, except in very special 
cases~\cite{sh2}. The first order correction to the circle is due to the spin, 
as the photon capture radius is different for co-rotating and counter-rotating 
particles. The boundary of the shadow has thus a dent on one side: the 
deformation is more pronounced for an observer on the equatorial plane 
(viewing angle $i = 90^\circ$) and decreases as the observer moves towards 
the spin axis, to completely disappear when $i = 0^\circ$ or $180^\circ$. 
The shape of the shadow can be characterized by the Hioki-Maeda distortion 
parameter~\cite{maeda}. If the inclination angle is known, the Hioki-Maeda
parameter only depends on the spin in the case of the Kerr background and
its measurement can thus be used to estimate $a/M$. However, if we
relax the Kerr BH hypothesis, the Hioki-Maeda parameter depends on both the spin 
and possible deviations from the Kerr solution. Eventually there is a degeneracy 
between these two quantities and therefore its measurement can only constrain an 
allowed region on the spin parameter--deformation parameter plane~\cite{z-sh}.

\subsection{Pulsar timing}

It is thought that a large population of pulsars can orbit near SgrA$^*$, and 
some of them may be in compact orbits to allow for precise measurements of 
the spacetime geometry around this object. Some deep pulsar searches have 
already been conducted~\cite{psearch}. Current 
results are consistent with the expectation of such a large population, but a 
pulsar sufficiently close to SgrA$^*$ to test general relativity has still to be found.

As discussed in Ref.~\cite{pulsar}, the possible observation of radio pulsars with
an orbital period shorter than about half year could provide a precise and clean 
measurement of the spin parameter of the central object from the Lense-Thirring 
effect. As the pulsar would still be far from the BH, such a technique would
really measure the spin parameter independently of the exact nature of the BH
candidate, just because possible deviations from the Kerr background 
correspond to higher-order terms in the multipole moment expansion of the
metric. In the case of a pulsar with an orbital period of one month or less, 
it would be also possible to measure the mass quadrupole moment of the
compact object and thus have a first test of the no-hair theorem~\cite{pulsar}, as 
in the case of a Kerr BH the mass quadrupole moment $Q$ is related to the 
mass $M$ and the spin angular momentum $J$ by the relation $Q = - J^2/M$. 
In what follows, it will be neglected such a possibility and it will be considered 
only the measurement of the spin. The measurement of $Q$ is indeed much 
more challenging, pulsar with a such short orbital period may be difficult to find,
and the approach can only test possible deviations from the Kerr quadrupole 
moment, while the measurements of the hot-spot period and the BH shadow 
are sensitive even to higher order corrections, as they depend on the spacetime
properties in the strong field regime.

It is worth noting that a similar measurement could be obtained 
from accurate astrometric data of ordinary stars orbiting at sub-mpc radii 
from SgrA$^*$, see e.g. Ref.~\cite{n-stars}. Even in this case, we would probe the 
weak gravitational field of the BH candidate and therefore we would measure 
the actual value of the spin from the frame-dragging precession, independently 
of the nature of the compact object. Like in the pulsar case, for very compact 
orbits it may be possible even to infer the mass-quadrupole moment. In the 
next section, I will talk about pulsar timing, but the same kind of constraints can 
be potentially obtained from astrometric observations, which have its own 
advantages and disadvantages.

\section{Simulations \label{s-3}}

The aim of this section is to consider some examples of possible future measurements 
with the three techniques discussed above and to study the constraints that can be 
obtained on the Cardoso-Pani-Rico deformation parameters. Of course, it is now 
impossible to predict the actual accuracy of these observations, so the final results
have to be seen as a general guide for future work, qualitatively correct but 
quantitatively to be taken with caution. While 
the first unconstrained deformation parameters are $\epsilon^t_3$ and 
$\epsilon^r_2$~\cite{cpr-m}, here the attention will be focused on $\epsilon^t_3$ and 
$\epsilon^r_3$, in order to consider the same order of deformation and quickly recover 
the Johannsen-Psaltis case when $\epsilon^t_3 = \epsilon^r_3$. The results and the
conclusions found for $\epsilon^t_3$ are qualitatively the same as the ones that would 
be obtained for a higher order $\epsilon^t_k$-type deformation parameter. The same is 
true for $\epsilon^r_3$ and higher order $\epsilon^t_k$-type deformation parameters. 
However, the more general possibility in which several deformation parameters can 
be non-vanishing at the same time is more complicated and requires a larger number 
of observations. The last example in this section considers the possibility of 
constraining $\epsilon^t_3$ and $\epsilon^r_3$ at the same time. In the following 
examples, it will be assumed that SgrA$^*$ has a mass of $4 \cdot 10^6$~$M_\odot$ 
and our viewing angle is $i = 60^\circ$.

It is well known that the Kerr metric can describe either BH (when 
$a/M \le 1$) or naked singularity ($a/M > 1$) solutions. The two scenarios can be 
experimentally distinguished by specific observational signatures~\cite{sh1,n-sing}.
In the case of the Cardoso-Pani-Rico metric, the situation is similar, but there are 
now three qualitatively different possibilities: BH solutions with a regular exterior, 
BH solutions with naked singularities, and naked singularity solutions. Like in 
the Kerr case, the three classes of spacetimes have their own observational 
signatures. An example is given in Ref.~\cite{n-dan}. In what follows, I will not 
look for similar observational signatures, which are beyond the aim of this work, 
and I will pay attention on the objects that may be interpreted as Kerr BHs.

\subsection{Example 1: slow-rotating Kerr BH}

\begin{figure*}
\begin{center}
\includegraphics[height=6cm]{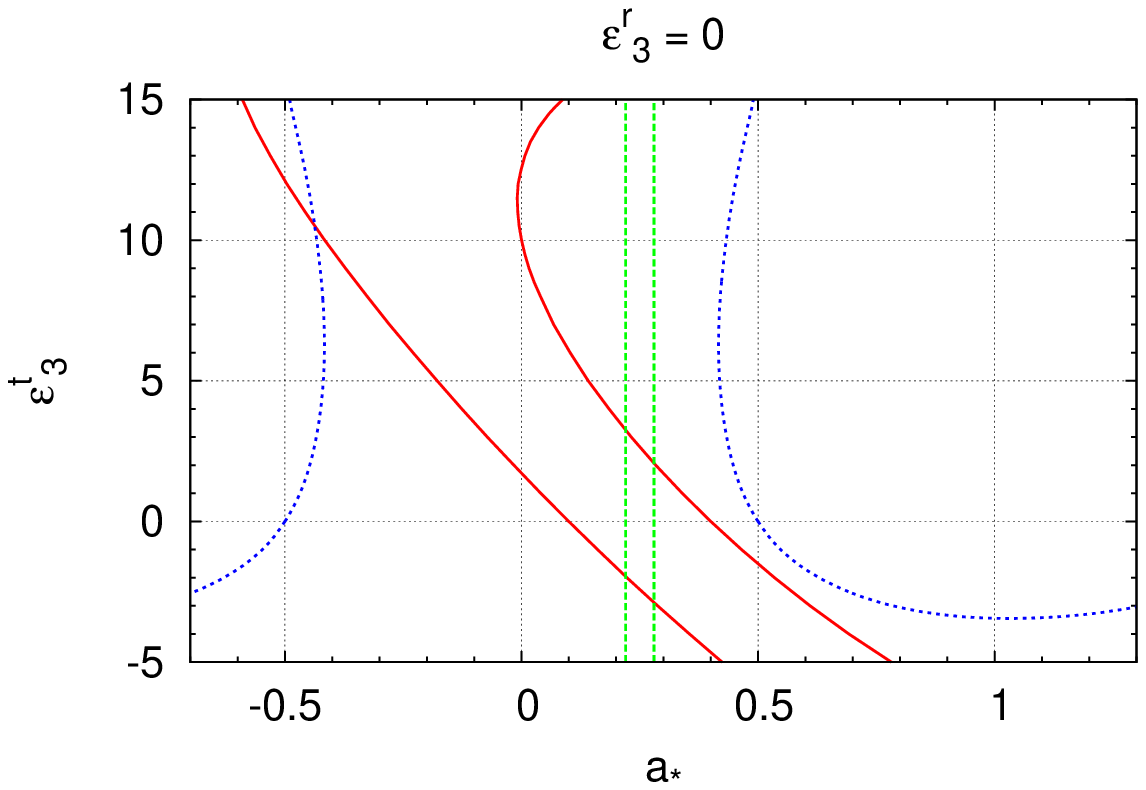}
\includegraphics[height=6cm]{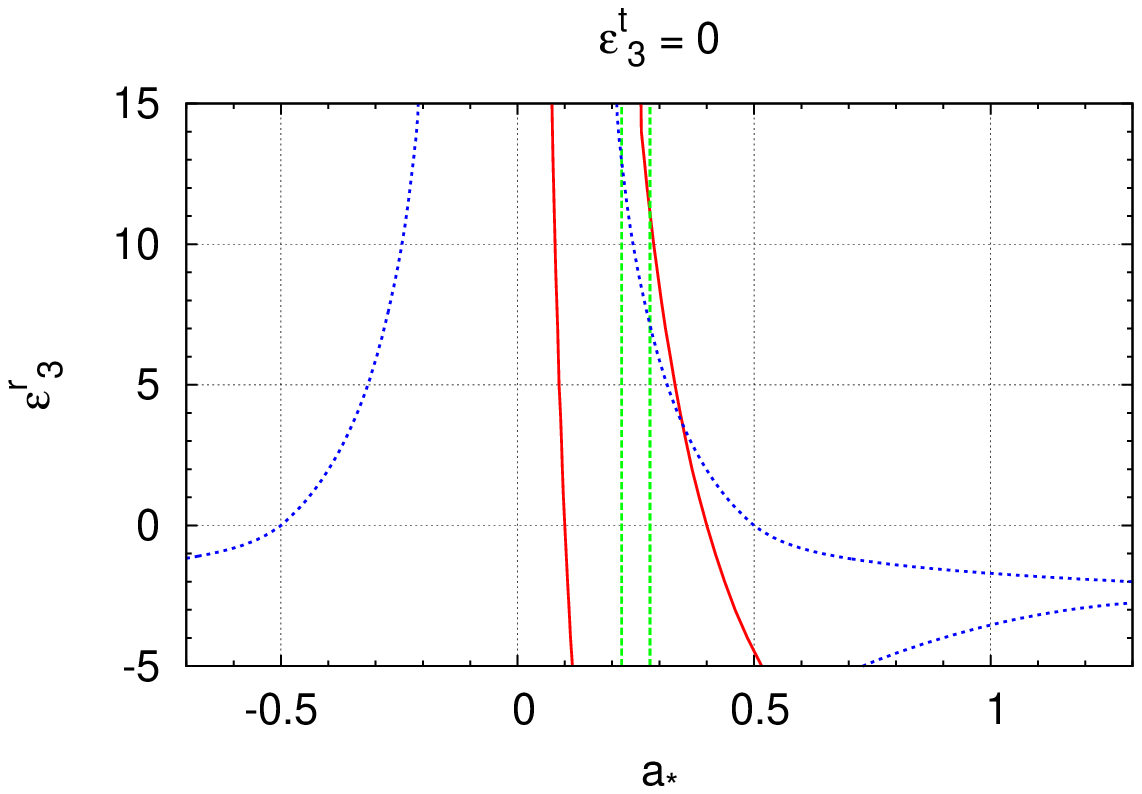}
\end{center}
\vspace{-0.5cm}
\caption{Possible constraints from the measurements of the ISCO frequency (red solid 
lines), of the Hioki-Maeda distortion parameter (blue dotted lines), and of the spin from 
a radio pulsar (green dashed lines). Left panel: constraints on the spin-$\epsilon_3^t$ 
plane assuming that all the other deformation parameters vanish. Right panel: constraints 
on the spin-$\epsilon_3^r$ plane assuming that all the other deformation parameters 
vanish. The constraints are obtained assuming ``reasonable'' future measurements of a 
Kerr BH with spin parameter $a_* = 0.25$. See the text for more details. \label{f1}}
\vspace{0.5cm}
\begin{center}
\includegraphics[height=6cm]{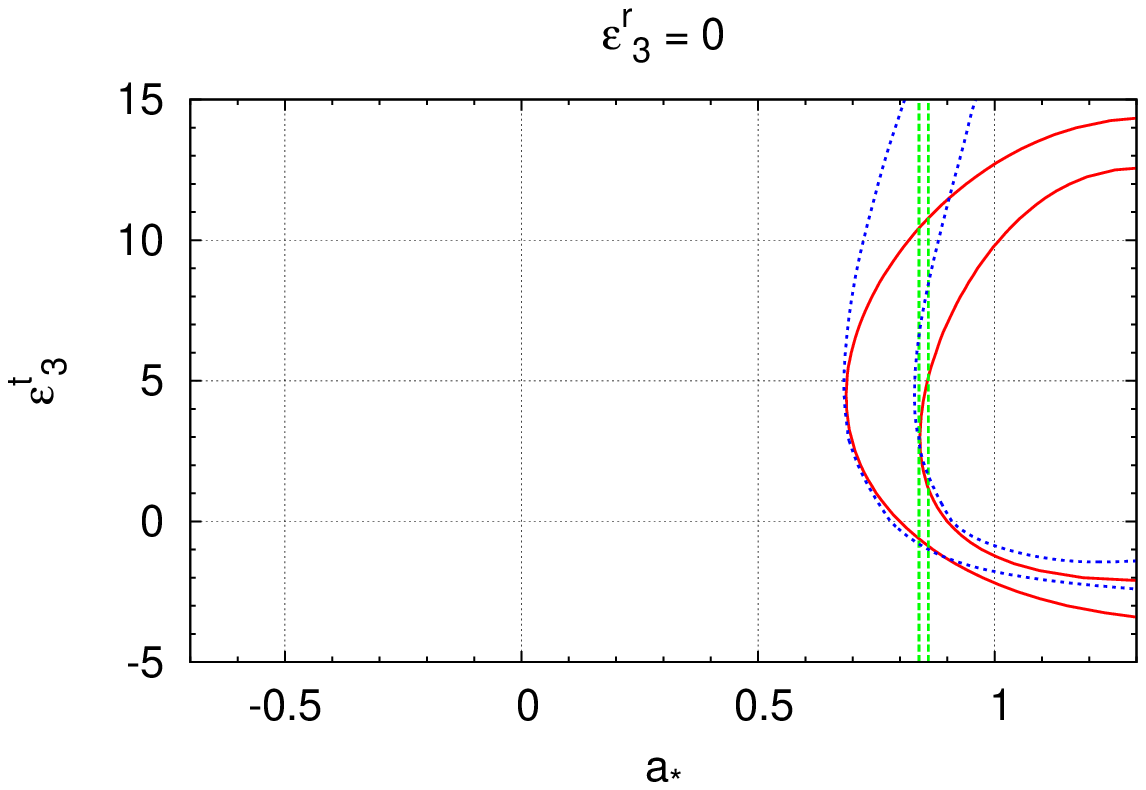}
\includegraphics[height=6cm]{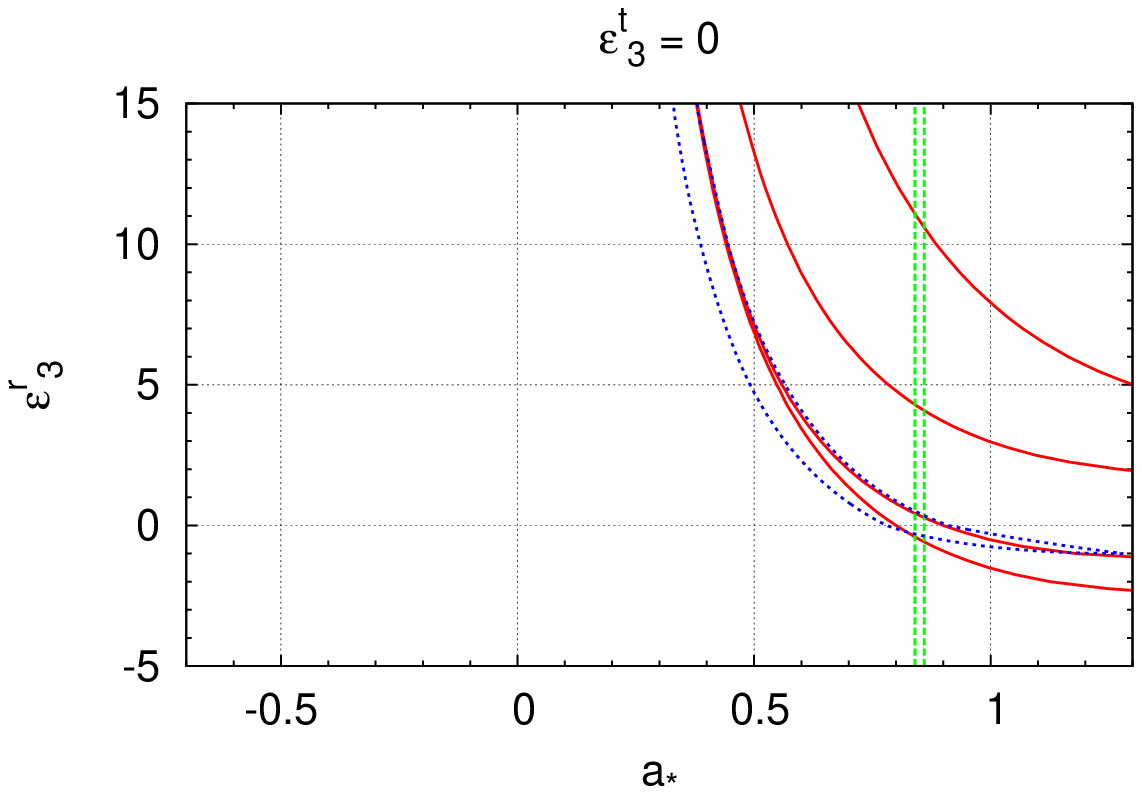}
\end{center}
\vspace{-0.5cm}
\caption{As in Fig.~\ref{f1}, with the constraints obtained assuming ``reasonable'' future
measurements of a Kerr BH with spin parameter $a_* = 0.85$. See the text for more 
details. \label{f2}}
\end{figure*}

As first example, let us consider the case in which SgrA$^*$ is a Kerr BH with 
dimensionless spin parameter $a_* = a/M = 0.25$. The observation of a blob of 
plasma orbiting close to SgrA$^*$ could provide an estimate of the Keplerian
orbital period at the ISCO radius. For a 4~million Solar mass Kerr BH with 
$a_* = 0.25$, $T_{\rm ISCO} = 25$~minutes. Let us suppose that the measurement 
is between 21.5 and 28.5~minutes, which would correspond to a spin estimate 
$0.10 < a_* < 0.40$ in the Kerr background. If we do not assume the Kerr 
background, such a measurement of $T_{\rm ISCO}$ provides an allowed 
region on the spin parameter--deformation parameter plane. Fig.~\ref{f1} shows 
this region (the area between the two red solid lines) in the case the only 
non-vanishing deformation parameter is $\epsilon^t_3$ (left panel) and in the 
case in which the only non-vanishing deformation parameter is $\epsilon^r_3$ 
(right panel). In the former case, there is a clear correlation between $\epsilon^t_3$ 
and $a_*$. In the latter case, the correlation is weak and actually the measurement 
of the hot spot orbital period is not really sensitive to the value of $\epsilon^r_3$.

To quantify the shape of the shadow, it is convenient to use the Hioki-Maeda distortion 
parameter $\delta$~\cite{maeda}. While future observations will use a more accurate 
approach, the Hioki-Maeda distortion parameter is a simple proxy to estimate the 
shadow departure from the shape of a circle. Let us also assume that our viewing angle 
is $i = 60^\circ$. For $i$ close to $0^\circ$, the boundary of the shadow would be 
necessarily a circle and its detection would provide no information about the nature
of SgrA$^*$. $i$ close to $90^\circ$ would be the most favorable position for a 
measurement, because the Hioki-Maeda distortion parameter would assume the 
highest value and any measurement would be more accurate. The measurement of 
$i$ can be independently obtained from the other two techniques. 
For a Kerr BH with $a_* = 0.25$ and
a viewing angle $i = 60^\circ$, $\delta = 0.006$, where $\delta = D/R$, $D$ is the ``dent''
and $R$ is the ``radius'' of the shadow (see Ref.~\cite{maeda} for the definitions).
Since our distance from the Galactic Center is about 8~kpc, the expected value of $R$ 
is around 25~$\mu$as. In this example, $D \sim 0.15$~$\mu$as, but such a measurement 
is out of reach. An optimistic accuracy on the measurement of the shape of the shadow 
is $\sim 0.5$~$\mu$as. It can thus make sense to consider a measurement 
$|\delta| < 0.023$. The latter corresponds to an estimate $|a_*| < 0.50$ in the Kerr 
background and to the areas between the blue dotted lines in Fig.~\ref{f1}. The 
constraints from the shadow will be surely weak in the case of a slow-rotating object. 
As one can see from the plots in the second paper in~\cite{sh2}, in non-Kerr metrics 
the Hioki-Maeda distortion parameter may be negative, so the sign of $\delta$ does 
not necessary fix the spin orientation.

Lastly, there is the pulsar measurement. Let us assume we get the measurement 
$0.22 < a_* < 0.28$. This would be the actual measurement, independently
of the spacetime geometry, because the pulsar orbit is at relatively large radii, where
possible deviations from the Kerr solution are more suppressed than the spin
term in the expansion of the background metric. In Fig.~\ref{f1}, the pulsar measurement
is the region between the two green dashed lines. The lines are vertical because 
the measurement is independent of the values of the deformation parameters.

\subsection{Example 2: fast-rotating Kerr BH}

Let us now assume that SgrA$^*$ is a Kerr BH with spin parameter $a_* = 0.85$. 
The result of the three measurements are shown in Fig.~\ref{f2}, respectively for 
$\epsilon^t_3$ (left panel) and $\epsilon^r_3$ (right panel), with all the other 
deformation parameters set to zero. The hot spot measurement is still represented by
the area between the two red solid lines. For a Kerr BH with $a_* = 0.85$, 
$T_{\rm ISCO} = 10.5$~minutes, and here it has been assumed an orbital period 
measurement between 9 and 12~minutes, which corresponds to a spin estimate
$0.80 < a_* < 0.90$ in Kerr. The constraint from the shadow is reported by the blue 
dotted lines. With the viewing angle $i = 60^\circ$, the Hioki-Maeda parameter
would be $\delta = 0.093$. Here, it has been assumed the measurement 
$0.071 < \delta < 0.120$. In the Kerr metric, it would correspond to a spin 
estimate $0.78 < a_* < 0.91$. Lastly, the observation from a radio pulsar could 
provide the bound $0.84 < a_* < 0.86$, and it is independent of the value of 
the deformation parameters.

\subsection{Example 3: constraining two deformation parameters}

\begin{figure*}
\begin{center}
\includegraphics[height=6cm]{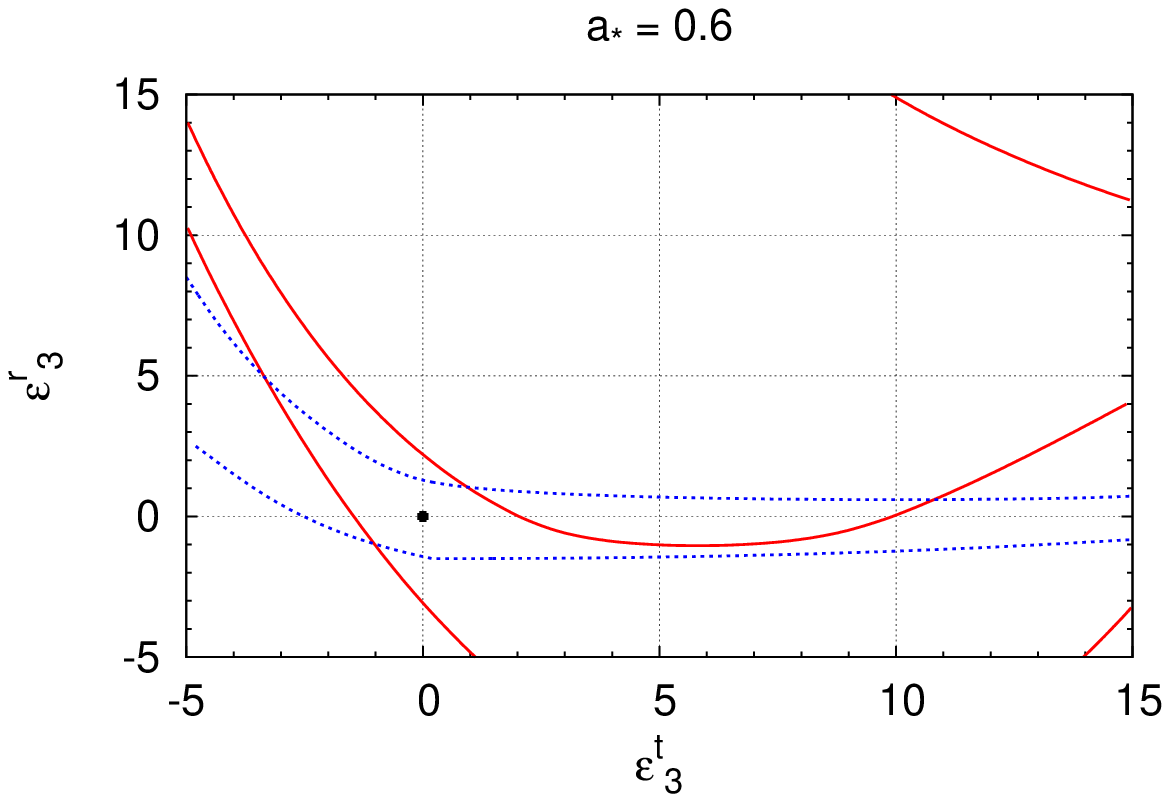}
\includegraphics[height=6cm]{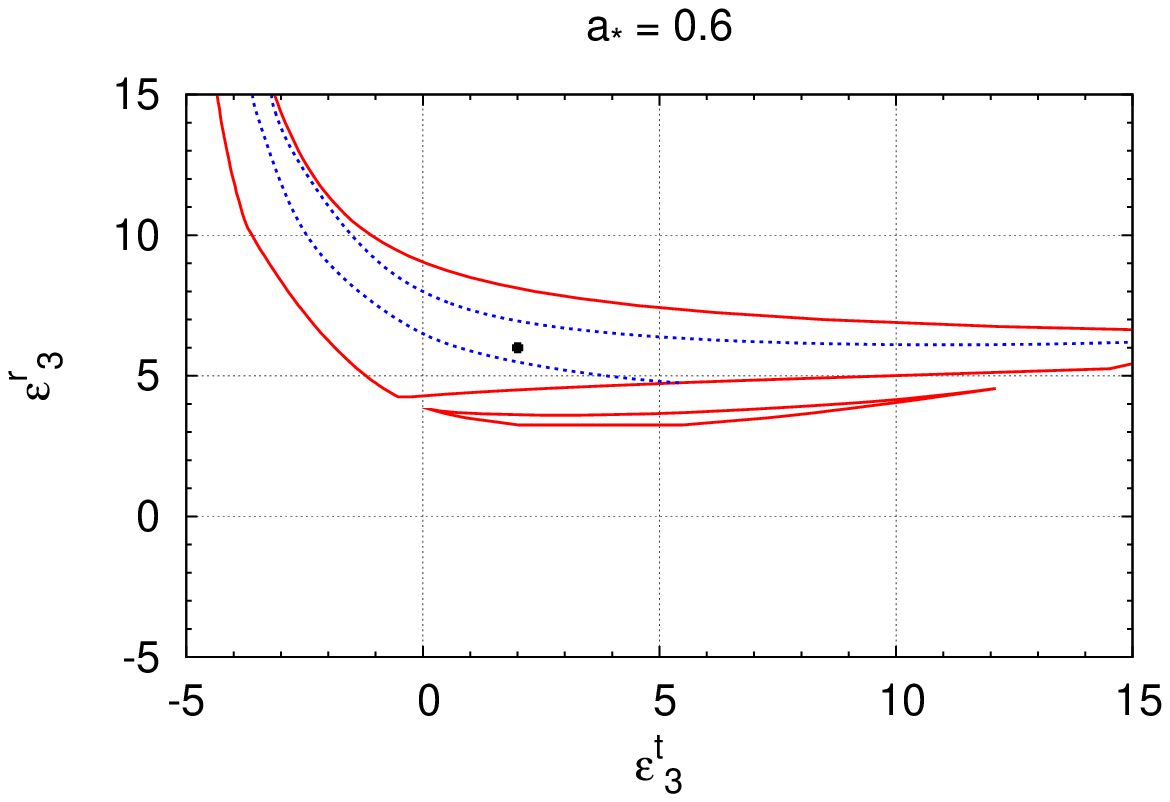}
\end{center}
\vspace{-0.5cm}
\caption{Constraints on the $\epsilon_3^t$-$\epsilon_3^r$ plane assuming to have 
an independent estimate of the spin $a_* = 0.6$ from a radio pulsar and that all the 
other deformation parameters vanish or are negligible. The red solid lines are for a 
measurement of the ISCO frequency. The blue dotted lines are for a measurement 
of the Hioki-Maeda distortion parameter. Left panel: constraints that may be obtained 
from a Kerr BH. Right panel: constraints that may be obtained from a non-Kerr BH 
with $\epsilon_3^t = 2$ and $\epsilon_3^r = 6$. The black dots indicate the reference 
BHs. See the text for more details. \label{f3}}
\end{figure*}

Lastly, I consider the case in which one wants to constrain $\epsilon^t_3$ and
$\epsilon^r_3$ at the same time. As the pulsar measurement can provide the 
correct value of the spin parameter independently of the exact nature of SgrA$^*$,
it is possible to fix the value of $a_*$ with the pulsar and study the constraints on the
$\epsilon^t_3$-$\epsilon^r_3$ plane. Let us assume that the pulsar observations
provide the estimate $a_* = 0.6$. We then consider the following two scenarios:
$i)$ SgrA$^*$ is a Kerr BH (left panel in Fig.~\ref{f3}), and $ii)$ SgrA$^*$ is a 
non-Kerr BH with $\epsilon^t_3 = 2$ and $\epsilon^r_3 = 6$ (right panel in 
Fig.~\ref{f3}).

In the former scenario of Kerr BH, for a 4~million Solar mass object the orbital period
at the ISCO is $T_{\rm ISCO} = 17$~minutes. Let us assume that observations
provide the measurement $14.5 < T < 19.5$~minutes, corresponding to a spin 
estimate $0.5 < a_* < 0.7$ in the Kerr background. The shadow of a Kerr BH with
spin $a_* = 0.60$ and seen with an inclination angle $i = 60^\circ$ has Hioki-Maeda
deformation parameter $\delta = 0.036$. Here it is assumed an uncertainty of 50\%,
so the measurement would be $0.018 < \delta < 0.054$ and the spin estimate in
the Kerr metric would be $0.45 < a_* < 0.70$. The hot spot and shadow measurements
are represented, respectively, by the red solid lines and blue dotted lines in Fig.~\ref{f3}.

In the scenario of non-Kerr BH, the ISCO orbital period would be $T_{\rm ISCO} = 
8.5$~minutes. Let us here assume a measurement $7.5 < T_{\rm ISCO} < 9.5$~minutes. 
The Hioki-Maeda distortion parameter would 
be $\delta = 0.219$. The measurement could be $0.200 < \delta < 0.240$. The final
constraints are shown in the right panel in Fig.~\ref{f3}.

\section{Concluding remarks \label{s-4}}

Astrophysical BH candidates are supposed to be the Kerr BHs of general relativity, 
but current observations cannot yet confirm this hypothesis. The key-point is that 
there is a strong correlation between the estimate of the spin and possible deviations 
from the Kerr solution. Current techniques to test the nature of these objects are 
the analysis of the thermal spectrum of thin accretion disks and of the shape of 
the iron K$\alpha$ line. However, they can only constrain a combination between 
the spin and the deformation parameters and a non-Kerr object could be interpreted 
as a Kerr BH with a different spin parameter. The combination of the measurements
from the disk's spectrum and the iron line is not very fruitful, as both the techniques 
are sensitive to the position of the ISCO radius.

If we want to test the Kerr BH hypothesis, it is necessary to combine several 
measurements of the same object and check whether the observational data 
require no deviations from the Kerr solution. For this reason, SgrA$^*$ may 
soon become the best object to test the actual nature of BH candidates. While 
there are currently no robust approaches to probe the spacetime geometry
around SgrA$^*$, there is the realistic possibility that, hopefully within a few 
years, a number of different techniques will be able to observe different 
relativistic effects from this source. The combination of these measurements 
can potentially allow to verify the nature of the supermassive 
BH candidate at the center of our Galaxy.

In this paper, I have focused the attention on three observations: the measurement 
of the ISCO frequency from the detection of a hot spot, the estimate of the 
distortion parameter of the shadow of SgrA$^*$, and the determination of the 
BH spin from accurate pulsar timing. All the three measurements can be potentially 
available within a few years. The hot spot and the shadow can test the geometry 
of the strong field regime, very close to the BH candidate. Pulsar measurements 
probe the weak field region at larger radii, where the perturbative regime holds. 
Possible constraints from future observations have been discussed within the 
Cardoso-Pani-Rico parametrization. In summary:
\begin{enumerate}
\item The discovery of a pulsar in a compact orbit can provide an accurate measurement 
of the spin parameter, independently of the exact nature of the compact object. 
\item In the case of a slow-rotating object, the measurement of the shadow can only 
provide very weak bounds. The combination of the hot spot and pulsar measurements 
can constrain the $\epsilon^t_k$-type deformation parameters, but it is much more 
difficult to do the same with the $\epsilon^r_k$-type deformation parameters. 
\item In the case of a fast-rotating object, the hot spot and the shadow can probably 
provide similar constraints, but their combination is anyway helpful to confirm the 
measurements, since it is challenging to have all the systematic effects perfectly under 
control. The combination of the three measurements can constrain the deformation 
parameter, and quite strong bounds can be obtained for the $\epsilon^r_k$-type 
deformation parameters. 
\item The possibility of constraining both the $\epsilon^t_k$ and $\epsilon^r_k$ 
deformation parameters at the same time is more challenging, and the three
techniques discussed in this paper may not be able to do it.
\end{enumerate}

{\it Acknowledgments ---}
This work was supported by the NSFC grant No.~11305038, 
the Shanghai Municipal Education Commission grant for Innovative 
Programs No.~14ZZ001, the Thousand Young Talents Program, 
and Fudan University.


\end{document}